\documentclass[aps,twocolumn,showpacs,preprintnumbers,amsmath,amssymb,nofootinbib,superscriptaddress,showkeys]{revtex4}

\usepackage{epsfig}
\usepackage{graphicx}
\usepackage{mathptmx}      % use Times fonts if available on your TeX system
\usepackage{latexsym}

\begin{document}

\title{$\gamma^* \gamma \rightarrow \pi^0$ transition form factor at low-energies from a model-independent approach}

\author{Pere Masjuan} \email{masjuan@ugr.es}
\affiliation{Departamento de F\'{i}sica Te\'{o}rica y del Cosmos and CAFPE,
Universidad de Granada, E-18071 Granada, Spain}

%\date{\today}

%\maketitle

\begin{abstract}

The recent measured $\gamma^*\gamma \rightarrow \pi^0$ transition form factor in the space-like region by the Belle Collaboration together with the previous published results by CLEO, CELLO and BABAR collaborations are analyzed using the mathematical theory of Pad\'{e} Approximants which provides a good and systematic description of the low energy region exemplified here with the extraction of the slope $a_{\pi}$ and curvature $b_{\pi}$ of the form factor in a model-independent way. The impact of them on the pion exchange contribution to the hadronic light-by-light scattering part of the anomalous magnetic moment $a_{\mu}$ is also discussed.

\end{abstract}

\pacs{12.38.-t, 12.38.Lg, 12.39.Fe,13.40.Gp}

\keywords{Pion Transition Form Factor, Pad\'{e} Approximants, Anomalous Magnetic Moment}

\maketitle

\section{Introduction}

The pion transition form factor (TFF) between a photon and a pion is extracted from the $e^+e^-\rightarrow e^+e^-\pi^0$ process where the $\pi^0$ is produced via the two-photon production mechanism. This transition is represented as a function of the photon virtualities as $F_{\pi^0\gamma^*\gamma^*}(q_1^2,q_2^2)$. The TFF is then extracted when one of the electrons is tagged. This electron emits a highly off-shell photon with momentum transfer $q_1^2\equiv-Q^2$ and is detected while the other, untagged, is scattered at a small angle and then its momentum transfer $q_2^2$ is near zero. The pion transition form factor is then defined as $F_{\pi^0\gamma^*\gamma^*}(-Q^2,0)\equiv F_{\pi^0\gamma^*\gamma}(Q^2)$.

The TFF was measured in the CELLO~\cite{CELLO} and CLEO~\cite{CLEO} experiments in the momentum transfer ranges $0.7-2.2$ GeV$^2$ and $1.6-8.0$ GeV$^2$, respectively. At 2009, the BABAR Collaboration extended these measurements in the $Q^2$ range from $4$ to $40$ GeV$^2$~\cite{BABAR}. And recently~\cite{Belle}, the Belle Collaboration has measured the form factor in the same BABAR's energy region with slightly different results on the high-energy region.

At low transferred momentum, the TFF can be described by the expansion:
\vspace{-0.2cm}
\begin{equation}
F_{\pi^0\gamma^*\gamma}(Q^2)\,=\, a_0\left(1+a_{\pi} \frac{Q^2}{m_{\pi}^2} + b_{\pi} \frac{Q^4}{m_{\pi}^4} + {\cal O}(Q^6)\right)\, ,
\end{equation}
\noindent
where the parameter $a_0$ can be determined from the axial anomaly~\cite{Adler:1969gk,Bell:1969ts} in the chiral limit of QCD, $a_0=\frac{1}{4\pi^2 f_{\pi}}$ with $f_{\pi}$ the pion decay constant.

The parameter $a_{\pi}$, the slope of the TFF, was measured by~\cite{Fonvieille:1989kj}, \cite{Farzanpay:1992pz}, and \cite{MeijerDrees:1992qb}, with the results $a_{\pi}=-0.11(3)(8)$, $a_{\pi}=0.026(24)(48)$, and $a_{\pi}=0.025(14)(26)$ respectively (the first error is statistic and the second systematic).
The CELLO Collaboration estimated $a_{\pi}$ to be $a_{\pi}=0.0326(26)_{stat}(26)_{sys}$ in Ref.~\cite{CELLO} using an extrapolation from the region of large space-like momentum transfer assuming a vector meson dominance (VMD) and using at zero transferred momentum the current experimental value for the partial decay width $\Gamma_{\pi^0\rightarrow\gamma\gamma}$ (which as we will see later is related to $F_{\pi^0\gamma^*\gamma}(Q^2=0)$, the axial anomaly).
The KTeV Collaboration also predicted $a_{\pi}=0.040(40)$ through a model-dependent fit to time-like data~\cite{Abouzaid:2008cd}.
The CELLO prediction however dominates the number quoted by the PDG \cite{PDG} since the direct measurements are less precise.

A VMD fit to all the available data (CELLO, CLEO, BABAR and Belle) would yield $a_{\pi}=0.0275(5)$ with a $\chi^2/d.o.f.=2.4$ ($d.o.f.$ meaning "degrees of freedom"), which means a 1.4 standard deviations from the CELLO result. This result suggests that the high-energy data may be important for determining low energy properties of the TFF.

One immediately comes to the question on how to improve on the quality of the fits to stabilize the predicted result and also on how to assign a systematic error to the fit procedure.

In Ref.~\cite{FFPades} it was suggested that the VMD is a first step on a sequence of particular rational approximations called Pad\'e Approximants (PA). In that reference was also suggested that using Pad\'e Approximants as a fitting functions to analyze the pion vector form factor in the space-like region, one can go beyond the VMD in a systematic approximation.

In the TFF case, this fact is of particular interest since the data from the BABAR collaboration cannot be easily accommodated in the VMD picture. With the help of these rational approximants, one could reach systematically the intermediate and high-energy experimental data producing, at the same time, an accurate results for the slope and curvature of the TFF at low-energies.

The Pad\'{e} technics provides with a simple, model-independent and systematic method of fitting data with a larger range of convergence than the simple polynomial fit or a VMD-like fit (such as the one used by CELLO collaboration to extract the $a_{\pi}$ parameter). Given a function $f(z)$ defined in the complex plane, the PA $P^N_M(z)$ are ratios of two polynomials $R_N(z)$ and $Q_M(z)$ (with degree $N$ and $M$ resp.) the coefficients of those exactly coincides with the coefficients of the Taylor expansion of $f(z)$ up to the highest order, i.e., $f(z)-P^N_M(z)={\cal O}(z^{N+M+1})$.

The PA method also provides with an estimation of a systematic error and could be also used to evaluate the impact of the vector excitations in the process considered. The technics described here were applied also on the search of resonance poles in Ref.~\cite{Montessus}.

There are several types of PA but as pointed out in Ref.~\cite{FFPades} the analytic properties of the function to be approximated determines what PA should be used. The time-like region is largely dominated by the $\rho-$meson contribution. The natural choice seems to be a $P^L_1(Q^2)$. On the other hand, since it is well known \cite{BL} that the TFF behaves like $1/Q^2$ at very large energies one could try to incorporate this information by considering a $P^N_{N+1}(Q^2)$ .

For explanatory reasons we show here how to constract a $P^L_1(Q^2)$ approximant [a $P^N_{N+1}(Q^2)$ is more involved and less illustrative].

Given a function $f(z)$ defined in the complex plane, a Pad\'e Approximant $P^L_1$ is defined~\cite{Baker}, without any loss of generality, by
\vspace{-0.2cm}
\begin{equation}\label{PAeq}
P_1^L(z,z_0)=\sum_{k=0}^{L-1}a_k(z-z_0)^k+\frac{a_L(z-z_0)^L}{1-\frac{a_{L+1}}{a_L}(z-z_0)}\, ,
\end{equation}
\noindent
where the coefficients $a_k$ are the Taylor coefficients of the corresponding $f(z)$ function that is been approximated.

Eq.~(\ref{PAeq}) shows that the pole $s_p$ of each $P^L_1$ is determined by the ratio $s_p=a_L/a_{L+1}$.

The TFF seems to be well described by the simple VMD ansatz. VMD relies on the accurate knowledge of the light meson spectra. When the spectral information is given in advance one should also take advantage of that information and consider other kinds of rational approximants. These are the Pad\'e-type approximants. In the Pad\'e-type approximants (PTAs) the poles of the Pad\'e are fixed to certain values (in our case, the resonances of the spectrum).

The simplest PTA sequence incorporates the lowest resonance, the $M_{\rho}$, and it is called $T^L_1$. The famous VMD ansatz is nothing but the simplest PTA, the $T^0_1$ approximant.

The purpose of this paper is twofold: first we want to extract the slope and the curvature of the Transition Form Factor using a sequence of $P^L_1$ approximants as fitting functions to the available experimental data. We demand an assignment of a systematic error to our predictions. Second we estimate the impact of our results on the light-by-light (LBL) contribution to the hadronic process on the muon $g-2$. We also comment in passing on the recent proposal to measure the transition form factor at low energy using the BESIII experiment.

We proceed as follows: In section \ref{sec:model} we study the reliability of the PA method as fitting functions to proceed then on section \ref{sec:fitreal} to analyze the real data. In section \ref{sec:g2} we consider the impact of the previous result on the LbyL contribution to the muon $g-2$. We finally collect all the results on the Conclusions section.

\section{Testing the method with a model}\label{sec:model}

Before applying the method to the experimental data to extract the slope and the curvature of the TFF, we want to test its reliability with a particular model. Since it has not been possible to describe rigourously the TFF from basic principals, several models have been developed during the last years with the purpose of analyzing the space-like data to extract fundamental QCD properties. In these Refs.~\cite{Arriola2006,Arriola2010,Mikhailov:2009kf,Dorokhov:2009dg,Radyushkin:2009zg,Polyakov:2009je,Li:2009pr,Noguera:2010fe,Roberts:2010rn,Agaev:2010aq,Klopot:2010ke,Kroll:2010bf,Pham:2011zi,Gorchtein:2011vf,Kampf:2011ty,Brodsky:2011yv,BrodskyLF,Bakulev:2011rp,Balakireva:2011wp,Czyz:2012nq,Lih:2012yu,Melikhov:2012bg,Bakulev:2012nh}  we try to summarize the large effort done on this purpose.

Considering this variety of models we examine instead of just one, three of them that we think are representative of the large amount of work done on this respect. Since our intent is to show the properties of our method, the selected models should describe well the experimental data but keep the complexity at a manageable level. This exercise will also provide a way to estimate the systematic error of our approximations.

For easy of reading we comment here about the first model and relegate the other two to the Appendix \ref{App}.

The first model considered is motivated by a Quark Model (e.g., \cite{Dorokhov:2009dg,Radyushkin:2009zg,Polyakov:2009je,Lih:2012yu}, see also~\cite{Pham:2011zi} for other $\log{(Q^2/M^2)}$ related models), although it can also be inspired by the lowest order perturbative QCD(pQCD) with a flat pion distribution amplitude (see, for example,~\cite{Radyushkin:2009zg}) or even by the BABAR fitting function~\cite{BABAR}. We named the model "log-model": %where a $1/Q^2$ is introduced to handle the model all the whole energy range:

\begin{equation}\label{Log-model}
F_{\pi^0\gamma^*\gamma}(Q^2)=\frac{M^2}{4 \pi^2 f_{\pi} Q^2}\log\left(1+\frac{Q^2}{M^2}\right)\, ,
\end{equation}
\noindent
with $M^2=0.6$GeV$^2$ and $f_{\pi}=92$MeV.

Expanding $F_{\pi^0\gamma^*\gamma}(Q^2)$ in Eq.~(\ref{Log-model}) in powers of $Q^2$ we obtain

\begin{equation}\label{TFF-exp}
F_{\pi^0\gamma^*\gamma}(Q^2)=a_0-a_1 Q^2 + a_2 Q^4 - a_3 Q^6+ {\cal O} (Q^8)\, ,
\end{equation}
\noindent
with known values for those $a_i$ coefficients (in particular $a_0=\frac{1}{4 \pi^2 f_{\pi}}$), as shown in the last column of Table~\ref{TableLog}.

\begin{table*}
\centering
\renewcommand{\arraystretch}{1.5}
\begin{tabular}{|c||c|c|c|c|c|c||c|}
\hline
 & $P^0_1$ &$P^1_1$ &$P^2_1$ & $P^3_1$ & $P^4_1$ & $P^5_1$ &  $F_{\pi^0\gamma^*\gamma}$ (exact) \\
\hline
$a_0 (GeV^{-1})$ & 0.2556 & 0.2694 &  0.2734 & 0.2746 & 0.2751 & 0.2752 &  0.2753 \\
$a_1 (GeV^{-3})$ & 0.1290 & 0.1716 &  0.1935 & 0.2051 & 0.2124 & 0.2166 &  0.2294 \\
$a_2 (GeV^{-5})$ & 0.0651 & 0.1147 &  0.1492 & 0.1725 & 0.1898 & 0.2013 &  0.2549 \\
\hline
$\sqrt{s_p} (GeV)$& 1.41 & 1.22  &  1.14  & 1.09  & 1.05  & 1.03  &  0.77\\
\hline
\end{tabular}
\caption{$a_0,a_1$ and $a_2$ low-energy coefficients of the log-model in Eq.~(\ref{Log-model}) fitted with a $P^L_1(Q^2)$
and its exact values (last column). We also include the prediction for the pole of each $P^L_1(Q^2)$ ($s_p$) to be compared with the lowest-lying meson in the model.}\label{TableLog}
\end{table*}

In order to illustrate the utility of the PA as a fitting functions we simulate the situation of the experimental data ~\cite{CELLO,CLEO,BABAR,Belle} with the model by considering the function Eq.~(\ref{Log-model}) evaluated at 22 points in the region $0.7\le Q^2 \le 5.5$ GeV$^2$, 16 points in the region $5.5\le Q^2 \le 12.5$ GeV$^2$ and 14 more points in the region $12.5\le Q^2 \le 35$ GeV$^2$. On top of these set of data points we add the value of $F_{\pi^0\gamma\gamma}(0,0)=\frac{1}{4\pi^2 f_{\pi}}$. All these data points have zero error because we want to obtain a pure systematic error on our fitting functions.

We construct a sequence of $P^L_1(Q^2)$ approximants with unknown coefficients as defined in Eq.~(\ref{PAeq}) and then we fit the set of data which yields a predictions for the $a_i$ coefficients. The results are shown in Table \ref{TableLog} where we go up to the $P^5_1$. The first $P^0_1$ has only two parameters ($a_0$ and $a_1$) and then $a_2$ is not a fitted but predicted through expansion. We also include on this table the position of the pole of each PA and the reader should notice how these poles, although showing a convergence pattern, differ from the lowest-lying vector mass used in the model of Eq.~(\ref{Log-model})\footnote{The model of Eq.~(\ref{Log-model}) has a branch cut starting at $Q^2\le-M^2$.}.

As expected~\cite{FFPades}, the sequence of PA converge to the exact result in a hierarchical way (much faster for $a_0$ than for $a_1$ and so on), achieving with the last PA $P^5_1$ a relative error of $0.04\%$, $5.6\%$ and $21.0\%$ for $a_0, a_1$ and $a_2$ respectively.

Similar results can be found by using as a fitting functions a sequence of PTAs as we said in the introduction. Thus, fixing the pole of the $T^L_1$ at $s_p= M^2=(0.77)^2$GeV$^2$, we obtain for the $T^5_1$ a relative error of $3\%$, $34\%$ and $92\%$ for $a_0, a_1$ and $a_2$ respectively. These results could be easily improved if instead of fixing the pole of our PTA on the starting point of the branch cut, i.e., at $s_p=(0.77)^2$GeV$^2$, we fix it at a different $s_p>(0.77)^2$GeV$^2$. For example, if $s_p=1$GeV$^2$ (value motivated by the result obtained with the previous $P^5_1$) the relative errors turn out to be $0.15\%$, $2.3\%$ and $14.4\%$ for $a_0, a_1$ and $a_2$ respectively. Since the PTA's prediction are very similar to PA's ones, we do not show explicitly the corresponding table. This simple exercise shows that fixing the pole of our approximant to the physical resonance, as in the VMD case, might not be the best strategy to follow for low-energy constant predictions, as extensively studied in Ref.~\cite{MasjuanThesis}.

The nice convergence pattern shown by our PA sequence should not be a surprise since it turns out that our model Eq.~(\ref{Log-model}) is a Stieltjes function and thus the convergence of the PA sequences is guaranteed by Pad\'e Theory~\cite{Stieltjes}.

On the other hand, it has been recently considered in Ref.~\cite{KLOE2} the possibility of KLOE-2 experiment at Frascati to measure the TFF at very low energies in the space-like region (for $0.01<Q^2<0.1 $GeV$^2$) and the width $\Gamma_{\pi^0\rightarrow \gamma\gamma}$ at the per cent level. This new low-energy data may reduce our systematic error for the PA $P^5_1$ to $4.2\%$ and $18\%$ for $a_1$ and $a_2$ respectively. An even better result might be obtained when the BES-III experiment at the $e^+e^-$ collider BEPC-II in Beijing will cover from the low-energy range up to CELLO energies, i.e., up to $Q^2\sim0.7$GeV$^2$ (which will turn out on systematic errors less than $3\%$ and $15\%$ for $a_1$ and $a_2$ resp. considering the feasibility study for BES-III performed in \cite{BESIII}). Indeed, the $\gamma\gamma$ physics program at BES-III for the measurement of pseudoscalar TFFs will allow to cover a wide $Q^2$ range below $10$GeV$^2$, the gap between KLOE-2 and CLEO experiments.

We analyze two more models in the Appendix \ref{App} using the same technic explained here and the similar results obtained with the three of them give us a confidence on our fit procedure. This exercise allows us to assign a systematic error for each element on the PA and the PTA sequences. To ascribe a particular (and conservative) systematic error and taking into account we do not know the structure of the whole TFF, we select the worse of the three cases as a guidance. For PA $P^5_1$, $5.6\%$, and $21\%$ and for PTA $T^5_1$, $5.4\%$, and $20\%$, as a relative systematic errors for $a_1$ and $a_2$ respectively.

\section{Fits to real data}\label{sec:fitreal}

With all the tools developed so far we can now proceed to analyze the real TFF. For this purpose we use all the available experimental data in the space-like region, which may be found in Refs. \cite{CELLO,CLEO,BABAR,Belle}, and also the recent measurement of the $ \Gamma_{\pi^0\rightarrow\gamma\gamma}$ decay width by the PrimEx Collaboration \cite{PrimEx}.

The form factor for real photons is related to the $\pi^0\rightarrow\gamma\gamma$ decay width:
\begin{equation}
F^2_{\pi^0\gamma\gamma}(q_1^2=0, q_2^2=0)=\frac{4}{\pi \alpha^2 m_{\pi}^3} \Gamma_{\pi^0\rightarrow\gamma\gamma}\, ,
\end{equation}
\noindent
with $\alpha=\alpha_{em}=1/137.0356$.

The experimental world average collected in the PDG tables \cite{PDG}  is $\Gamma^{PDG}_{\pi^0\rightarrow\gamma\gamma}=7.74 \pm 0.48$ eV, although we use here the PrimEx Collaboration result \cite{PrimEx}, which using a Primakoff effect experiment at JLab, has improved significantly the accuracy, reporting the value $\Gamma_{\pi^0\rightarrow\gamma\gamma}=7.82 \pm 0.14 \pm 0.17$ eV.

\begin{figure*}
  \includegraphics[width=4.5in]{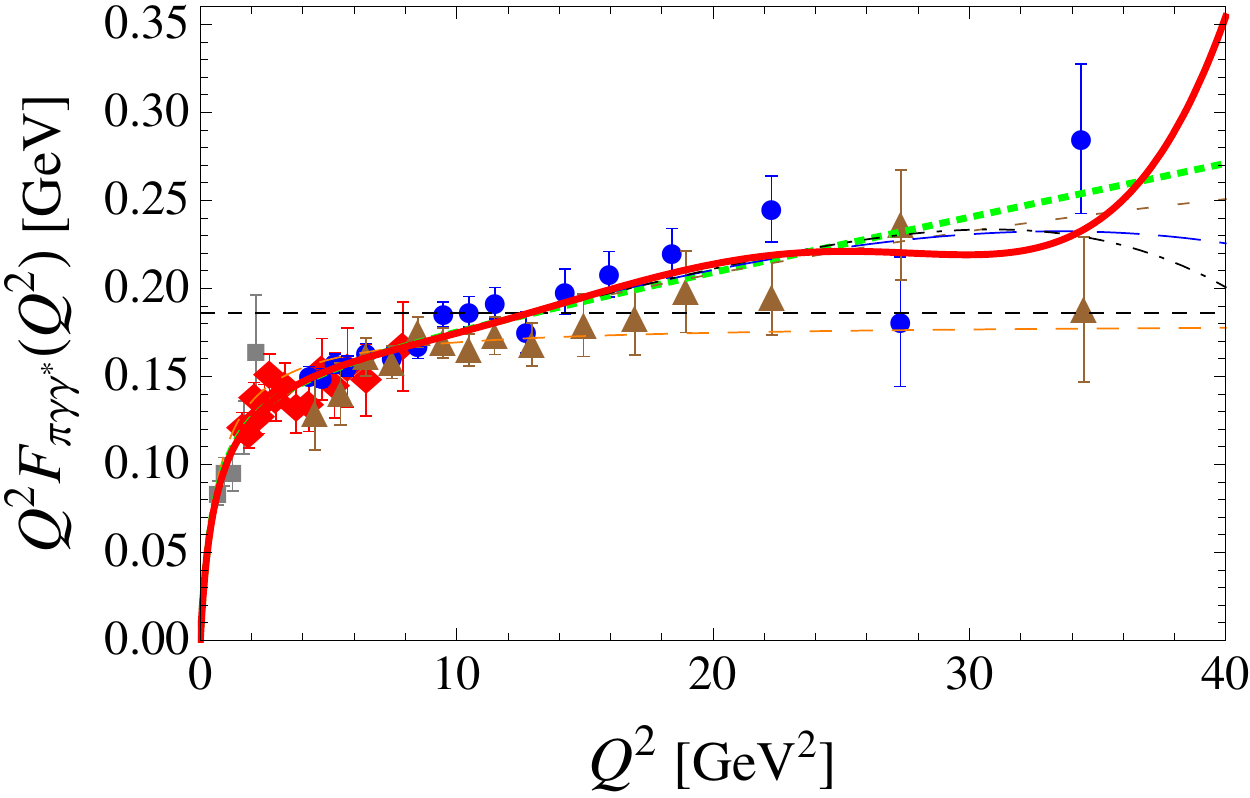}\\
\caption{The $P^L_1$ sequence compared with the $\gamma^* \gamma \rightarrow \pi^0$ Transition Form Factor data from CELLO (grey squares)~\cite{CELLO}, CLEO (red diamonds)~\cite{CLEO}, BABAR (blue circles)~\cite{BABAR} and Belle (brown triangles)~\cite{Belle}: $P^0_1$ (orange dashed), $P^1_1$ (green dotted), $P^2_1$ (brown short-dashed), $P^3_1$ (blue long-dashed), $P^4_1$ (black dot-dashed) and $P^5_1$ (red solid). Black dashed line indicates the pQCD result. }\label{plot1}
\end{figure*}

\subsection{Fits with the rational approximants}

The fits with the $P^L_1$ sequence to the space-like data points in Refs.~\cite{CELLO,CLEO,BABAR,Belle} determine those $a_k$ coefficients that best interpolate them. As always, when fitting experimental data one should find a compromise between the increase of fit errors and decrease of systematic ones when increasing the order $L$ of the $P^L_1$. Figure \ref{plot1} shows the experimental data obtained by CELLO (grey squares)\footnote{CELLO data points $D_i$ are extracted from Ref.~\cite{CELLO} using the following normalization:  $D_i= \left(\frac{64\pi N_i}{(4\pi\alpha)^2m_{\pi}^3}\right)^{1/2}$, with the $N_i=\frac{F^2(Q_i^2)m_{\pi}^3}{64\pi}$ provided in that reference and $\alpha=1/137.036$.}, CLEO (red diamonds), BABAR (blue circles) and Belle (brown triangles) together with the pQCD prediction (horizontal black dashed line). The red curve on Fig.\ref{plot1} is our best approximant, the $P^5_1$.

\begin{figure*}
  \includegraphics[width=3in]{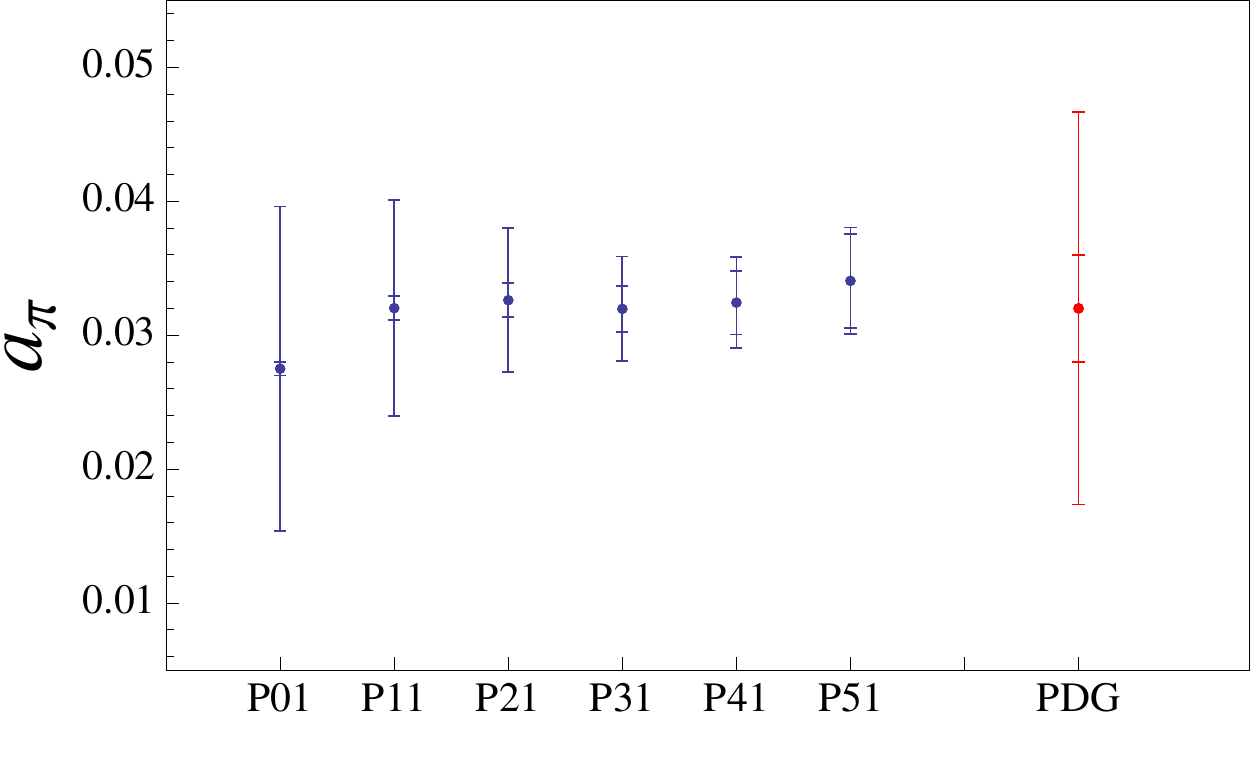}
  \includegraphics[width=3in]{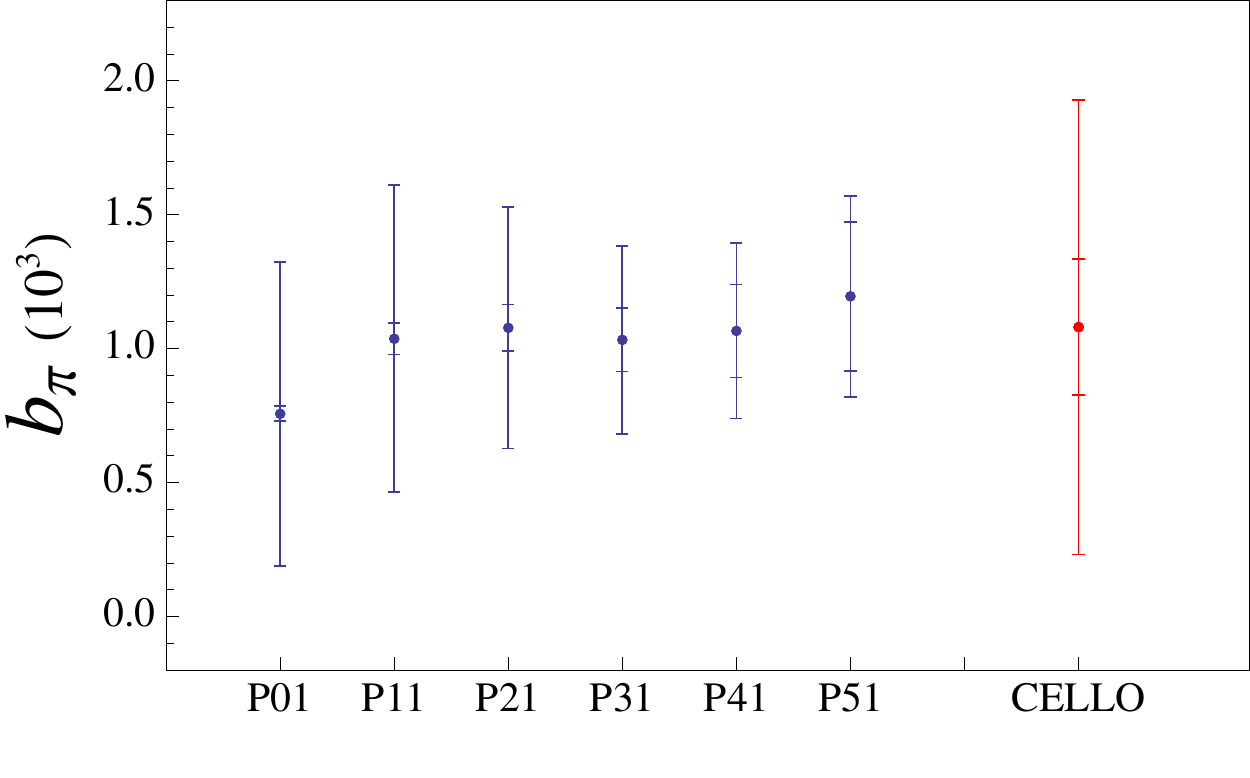}\\
\caption{$a_{\pi}$ (left) and $b_{\pi}$(right) predictions with the $P^L_1$ up to $L=5$. The internal band is the statistical error from the fit and the external one is the combination of statistical and systematic errors determined in the previous section. }\label{ApiPA}
\end{figure*}

In Fig.~\ref{ApiPA} we show the results for the prediction of the slope and curvature parameters $a_{\pi}$ and $b_{\pi}$ with the $P^L_1$ up to $L=5$. Approximants with $L>5$ have the new coefficients compatible with zero and then do not introduce new information with respect to $P^5_1$. The internal errors shown in Fig.\ref{ApiPA} are only statistical and the external ones are a quadratic combination of statistical and systematic errors, the latter determined in the previous section. For completeness we also ascribe a $45\%$ of systematic error to the PDG slope value\footnote{Again, this systematic error is obtained comparing the VMD result with the exact one in Tables \ref{TableLog},\ref{TableRegge}, and \ref{TableLF}.}. The curvature parameters has never been measured so for easy of comparison we expand the VMD fit used by the CELLO collaboration up to that order with the corresponding systematic error.

As expected from the models studied, we see in these figures a nice convergence pattern for both $a_{\pi}$ and $b_{\pi}$.

\begin{figure}
  \includegraphics[width=3in]{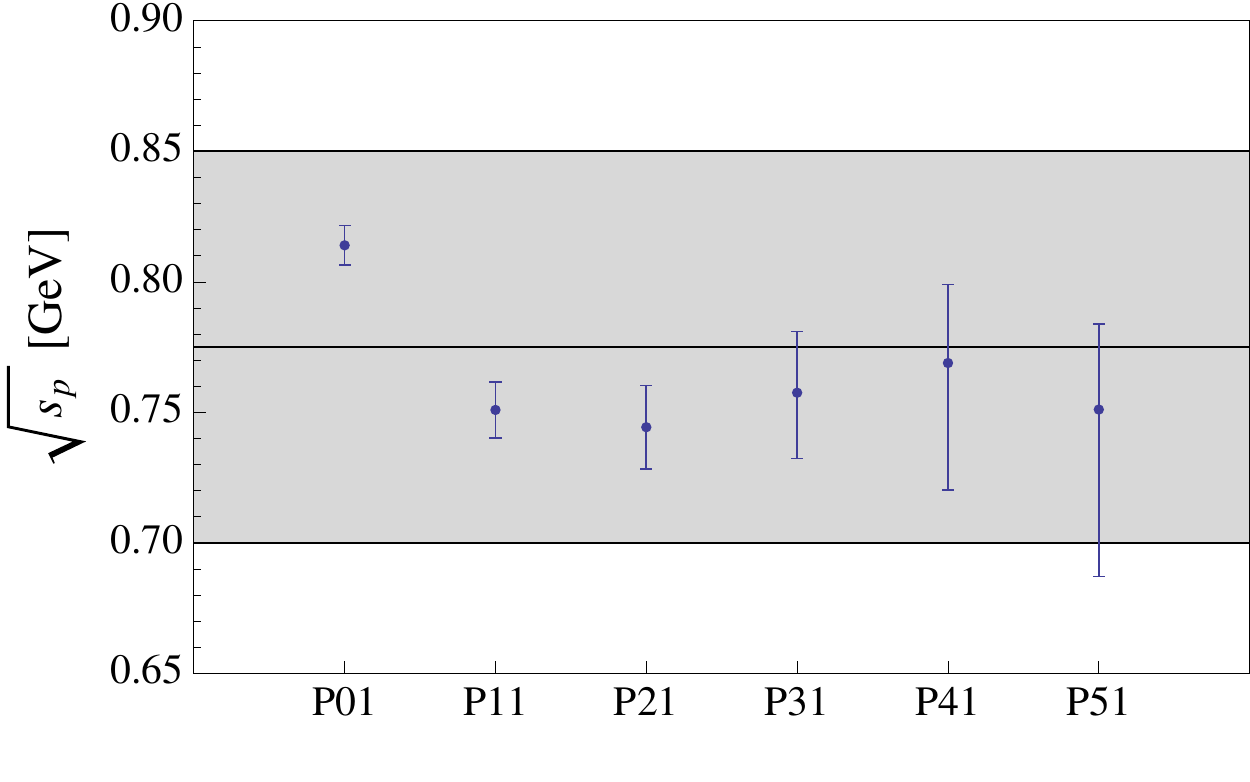}\\
\caption{Position of the pole $\sqrt{s_p}$ for the different $P^L_1$. For comparison we also show (gray band) the range $M_{\rho}\pm\Gamma_{\rho}/2$ corresponding to the physical $\rho-$meson value.}\label{fig:sp}
\end{figure}

The PA $P^5_1$ yields
\begin{equation}\label{res:P51a}
a_{\pi}=0.0340(35)_{\mathrm{stat}}(19)_{\mathrm{sys}}\, ,
\end{equation}
and
\begin{equation}\label{res:P51b}
b_{\pi}=1.20(28)_{\mathrm{stat}}(25)_{\mathrm{sys}}\times 10^{-3}\, ,
\end{equation}
\noindent
with a $\chi^2/d.o.f.=0.80$, where the systematic error is estimated from the previous section ($5.6\%$ for $a_{\pi}$ and $21\%$ for $b_{\pi}$). We also extract the position of the PA pole $s_p=a_L/a_{L+1}$. This ratio is shown in Fig.\ref{fig:sp} together with a band corresponding to the physical value $M_{\rho}\pm \Gamma_{\rho}/2$ where $M_{\rho}=0.7755$GeV and $\Gamma_{\rho}=0.155$GeV is believed to be the dominant resonance contribution. For the $P^5_1$ the pole is located at $\sqrt{s_p}=0.75^{+0.03}_{-0.06}$GeV, well in this band.

It is interesting to notice the slightly larger results for the slope obtained with $P^L_1$ with $L>1$, manifesting the need of a systematic procedure for going beyond VMD. It turns out that the larger $L$, larger the sensibility of the $P^L_1$ to the high-energy data. In this respect, the recent Belle data is crucial to obtain an accurate low-energy prediction since up to now BABAR data was dominating the high-energy region (see, for example, Ref.~\cite{Montessus} for a preliminary study without Belle data).

Our final result is a fit to all the available data but for a deeper understanding of PA as a fitting functions for high-energy data we consider two different scenarios where first no Belle data is considered and second no BABAR data is considered (Table \ref{Tablecomparison}). Surprisingly enough, the results shown in Table \ref{Tablecomparison} (where \textit{All} stands for all the available data) are nicely compatible within errors although with slightly different central values. All the results are obtained with a $P^5_1$.

\begin{table}
\centering
\renewcommand{\arraystretch}{1.5}
\begin{tabular}{|c||c|c|c|c|}
\hline
 Data 			& $a_{\pi}$ 	&$(10^3)b_{\pi}$ &$\sqrt{s_p}$ 		& $\chi^2/d.o.f.$  \\
\hline
All 			& $0.0340(35)$	& $1.20(28)$	 &  $0.75_{-0.06}^{+0.03}$ 	& 0.80  \\
CELLO+CLEO+BABAR 	& $0.0348(39)$	& $1.26(32)$	 &  $0.73_{-0.06}^{+0.04}$ 	& 0.61  \\
CELLO+CLEO+Belle 	& $0.0326(39)$	& $1.08(30)$	 &  $0.76_{-0.06}^{+0.05}$ 	& 0.49  \\
\hline
\end{tabular}
\caption{Slope and curvature of the TFF predictions with a $P^5_1$ with different sets of data.}\label{Tablecomparison}
\end{table}

Finally, we want to apply a last test of robustness to the method which is fits to subsets of data should return compatible results, i.e., fitting data up to $10$GeV$^2$, up to $20$GeV$^2$ and up to $36$GeV$^2$ (all the data) should be the same (unless there are unknown problems with the data like normalization or systematics). The results are shown in Table \ref{Tablesets} where we also indicate the best $P^L_1$ that fits the particular subset of data. These results are nicely compatibles, otherwise we should take the difference as a new source of systematic error.

\begin{table}
\centering
\renewcommand{\arraystretch}{1.5}
\begin{tabular}{|c||c|c|c|}
\hline
  			& best PA	& $a_{\pi}$ 	 & $\chi^2/d.o.f.$  \\
\hline
Data up to 10GeV$^2$	& $P^3_1$	& $0.0364(51)$	 & 0.53  \\
Data up to 20GeV$^2$	& $P^4_1$	& $0.0327(35)$	 & 0.69  \\
All			& $P^5_1$	& $0.0340(35)$	 & 0.80  \\
\hline
\end{tabular}
\caption{Slope of the TFF prediction with with different sets of data as described in the main text.}\label{Tablesets}
\end{table}

\begin{figure}
  \includegraphics[width=3in]{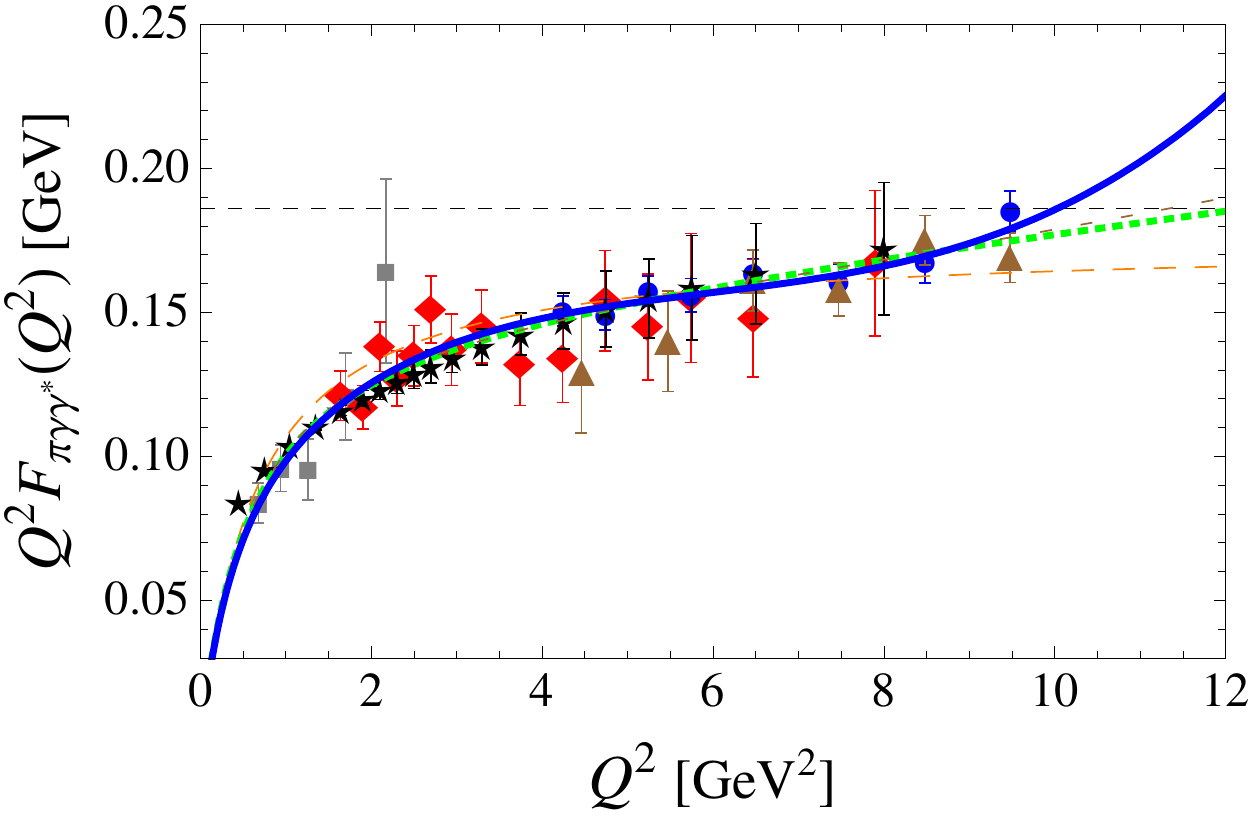}\\
\caption{The $P^L_1$ sequence compared with the $\gamma^* \gamma \rightarrow \pi^0$ Transition Form Factor data up to $10$GeV$^2$ \cite{CELLO,CLEO,BABAR,Belle,BESIII}: $P^0_1$ (orange dashed), $P^1_1$ (green dotted), $P^2_1$ (brown short-dashed), $P^3_1$ (blue solid). Black dashed line indicates the pQCD result. }\label{plot2}
\end{figure}

For illustrative purpose we show on Fig.~\ref{plot2} the result when fitting data up to $10$GeV$^2$. In this case we include also the feasibility study for BES-III experiment performed in Ref.~\cite{BESIII}. With a $P^3_1$ we obtain \footnote{PA with larger $L$ do not introduce new information. BES-III data will be crucial to improve on this result.} $a_{\pi}=0.036(6)$ and $b_{\pi}=1.41(65)$ where the errors are statistical and systematical with $\chi^2/d.o.f.=0.53$, to be compared with the results on Table \ref{Tablecomparison}. In this scenario the pole of the $P^3_1$ is located at $\sqrt{s_p}=0.73_{-0.05}^{+0.09}$GeV.

\subsection{Other Pad\'e Approximants}

\subsubsection{$P^L_2$ Pad\'e Approximants}

The experimental data so far considered range up to $36 $GeV$^2$ then a natural extension of the previous analysis would include higher resonances although the form factor is believed to be dominated by the $\rho(770)$ meson. In such a way, the consideration of two-pole $P^L_2$ will give us a way to asses any possible systematic bias in our $P^L_1$ analysis.

In this case our best approximant is the $P^3_2$.  This approximant yields
\begin{equation}\label{res:P32}
a_{\pi}=0.0324(20)\quad \mathrm{and}\quad b_{\pi}=1.07(15)\times 10^{-3} \, ,
\end{equation}
\noindent
with a $\chi^2/d.o.f.=0.71$, nicely compatible with our previous determination in Eqs.~(\ref{res:P51a}) and (\ref{res:P51b}).
Despite this result, the poles of that approximant are located at $s_{p1}=0.53(6)- i 0.01(1)$ and $s_{p2}=0.56(2)+i 0.01(1)$, where we can see a certain parameter space region where the poles may became eventually complex-conjugated\footnote{Complex-conjugate to render the approximant real.}.

\subsubsection{$T^L_1$ Pad\'e-Type Approximants}

On the other hand, since the value of the physical $\rho-$meson mass is well known, it is natural to attempt to include this information in our analysis through the PTAs. We have seen on the previous section, however, that locating the pole of a PTA exactly at the physical counterpart is not the best strategy. In fact, we learnt that $s_p>M_{\rho}^2$ but it is not clear what particular value we should use. To evaluate a possible systematic error on this choice, we range the PTA pole in between the band drawn by the PA results in Fig.\ref{fig:sp}, i.e, $\sqrt{s_p}=0.73-0.83$GeV.

With a $T^L_1$ sequence we go up to $T^5_1$ and obtain also a nice and smooth convergence pattern for both $a_{\pi}$ and $b_{\pi}$ parameters. With our best PTA, we obtain
\begin{equation}\label{res:T51}
a_{\pi}=0.0302(28) \quad \mathrm{and} \quad  b_{\pi}=0.92(18)\times10^{-3}\, ,
\end{equation}
\noindent
with a $\chi^2/d.o.f.=0.78-0.87$, where the errors are mainly the systematics of the pole range.

\subsubsection{$P^N_{N+1}$ Pad\'e Approximants}

As suggested in the Introduction, we may attempt to include the asymptotic behavior of the TFF~\cite{BL} in our fits by considering a $P^N_{N+1}$ sequence. With these approximants we can go up to the $P^2_3$ which yields
\vspace{-0.2cm}
\begin{equation}\label{res:P23}
a_{\pi}=0.0331(45) \quad \mathrm{and} \quad  b_{\pi}=1.11(27)\times10^{-3}\, ,
\end{equation}
\noindent
with a $\chi^2/d.o.f.=0.73$.

The $P^2_3$ approximant has the right fall-off as $Q^{-2}$ but the corresponding coefficient (which reads $0.17\pm1.8 $GeV) is not correctly predicted due to its large statistical error. This asymptotic coefficient is known to be $2 f_{\pi}$ by first principles~\cite{BL}. It seems logical to trying to include this information on the $P^2_3$. Using the asymptotic coefficient when constructing the $P^2_3$, we obtain a constrained approximant called $P'^2_3$ which after fitting the TFF data yields, for the low-energy coefficients,
\vspace{-0.2cm}
\begin{equation}\label{res:Pp23}
a_{\pi}=0.0332(25) \quad \mathrm{and} \quad  b_{\pi}=1.13(19)\times10^{-3}\, ,
\end{equation}
\noindent
with a $\chi^2/d.o.f.=0.70$.

It is remarkable that these results with the $P'^2_3$, which makes use of all the experimental data and the asymptotic limit at once, are nicely compatible with all the previous results. This approximant seems to suggest that the scale where the pQCD should be applied is much further away than the last BABAR and Belle data points.

\subsection{Final Result}

The results shown in Eqs.~(\ref{res:P51a}-\ref{res:Pp23}) agree quite well and combining them our final weighted average result yields:

\begin{equation}\label{FinalResult1}
a_{\pi}=0.0324(12)_{stat}(19)_{sys}\, ,
\end{equation}
\noindent
and
\begin{equation}\label{FinalResult2}
b_{\pi}=1.06(9)_{stat}(25)_{sys}\times10^{-3}\, ,
\end{equation}
\noindent
to be compared with other theoretical determinations: from a Regge analysis, $a_{\pi}=0.032(1)$ \cite{Arriola2010}; from ChPT at the loop level with $\mu=M_{\rho}$, $a_{\pi}=0.036$ \cite{Bijnens:1989jb}; from  a study of the Dalitz decay $\pi^0\rightarrow e^+e^-\gamma$, $a_{\pi}=0.029(5)$ \cite{Kampf:2005tz}; from a hard-wall holographic models of QCD, $a_{\pi} \approx 0.031$ \cite{Grigoryan:2008up} and $a_{\pi} \approx 0.035$ \cite{Grigoryan:2008cc}; from a soft-wall holographic model of QCD, $a_{\pi}=0.024(5)$ \cite{Colangelo:2011xk}\footnote{This number is obtained through the large-$N_c$ limit relation $C_{22}^W=\frac{a_{\pi}N_c}{64 \pi^2 m_{\pi}^2}$ \cite{Kampf:2005tz} with $C_{22}^W=6.3\times10^{-3}$ obtained in \cite{Colangelo:2011xk}. Indeed, with our final value for $a_{\pi}$ we predict $C_{22}^W=8.4(9)\times 10^{-3}$GeV$^{-2}$.}; and finally from the compilation of holographic models on Ref.\cite{Oscarg2}, $a_{\pi}=0.031(6)$, where the error is estimated by the spread of the different results obtained from these models.

In the next section we explore possible consequences of our final results in Eqs.~(\ref{FinalResult1}) and (\ref{FinalResult2}) on the light-by-light scattering contribution to the anomalous magnetic moment of the muon.

\section{Implications on the hadronic Light-by-Light contribution to the $(g-2)_{\mu}$}\label{sec:g2}

We can use the results in Eqs.~(\ref{FinalResult1}) and (\ref{FinalResult2}) to constrain any model that estimates the pion-exchange piece to the Light-by-Light scattering contribution to the $(g-2)_{\mu}$, the $a_{\mu}^{LbyL;\pi^0}$ term. As an example, we consider the so called $LMD+V$ model (defined in Ref.~\cite{KcnechtNyffeler}) to account for that contribution:

{\small
\begin{align}\label{LMDV}
&F_{\pi^0\gamma^*\gamma^*}^{LMD+V}(Q_1^2,Q_2^2)=\nonumber\\
&\frac{f_{\pi}}{3} \frac{-Q_1^2Q_2^2(Q_1^2+Q_2^2)+h_1(Q_1^2+Q_2^2)^2+h_2 Q_1^2Q_2^2-h_5(Q_1^2+Q_2^2)+h_7}{(Q_1^2+M^2_{V_1})(Q_1^2+M^2_{V_2})(Q_2^2+M^2_{V_1})(Q_2^2+M^2_{V_2})}\, .
\end{align}
}

The TFF is related to the LMD+V model Eq.~(\ref{LMDV}) when one of the photons on the latter is on-shell. That means we cannot fix all the free parameters ($h_i$, with $i=1,2,5,7$, and $M_{V_1}, M_{V_2}$) on this LMD+V model at once. We need more information, for example, from the high-energy region ($Q^2 F_{\pi^0\gamma^*\gamma}(Q^2,0)=2 f_{\pi}$, Ref.~\cite{BL}). If we match the high-energy limit, we find $h_1=0$ and $h_5=-6 M_{V_1}^2M_{V_2}^2$. The axial anomaly on the low-energy limit fixes  $h_7=-\frac{N_c}{4\pi^2f_{\pi}^2}M_{V_1}^4M_{V_2}^4$. With these results and $h_2=0$ as suggested in Ref.~\cite{KcnechtNyffeler}, we can use the slope and the curvature of the TFF to fix $M_{V_1}$ and $M_{V_2}$. We find $M_{V_1}^2=0.33(11)$GeV$^2$ and  $M_{V_2}^2=0.94^{+0.99}_{-0.25}$GeV$^2$ and we obtain $a_{\mu}^{LbyL;\pi^0}=5.4(5)\times 10^{-10}$.

\section{Conclusions}

In this paper, we analyzed the collection of all the experimental data on the $\pi^0\gamma^*\gamma$ Transition Form Factor at low-energies with a model-independent approach based on Pad\'e Approximants and we obtain the slope $a_{\pi}=0.0324(12)_{stat}(19)_{sys}$ and curvature $b_{\pi}=1.06(9)_{stat}(25)_{sys}\times 10^{-3}$ of the form factor. The method is simple and systematic and provides a model-independent estimation of all the systematic errors. We analyzed the impact at low-energy of the Belle and BABAR high-energy data and also the future BES-III data.
We also evaluate the implications of these results on the pion-exchange contribution on the light-by-light scattering part of the anomalous magnetic moment of the muon. Using the well-known LMD+V parametrization and the Pad\'e Theory technics we estimate that contribution to be $a_{\mu}^{LbyL;\pi^0}=5.4(5)\times 10^{-10}$.

\vspace{1cm}

{\bf Acknowledgements}
\\

We thank F. Cornet, R. Escribano, A. Nyffeler, E.R.Arriola, and V.Savinov for discussions, B. Kloss, E.Prencipe, and M. Vanderhaeghen for providing us with the feasibility study at BES-III experiment and also F. Cornet and E.R.Arriola for a critical reading of the manuscript. This work has been supported by MICINN, Spain (FPA2006-05294), the Spanish Consolider-Ingenio 2010 Programme CPAN (CSD2007-00042) and by Junta de Andaluc\'ia (Grants P07-FQM 03048 and P08-FQM 101).

\appendix
\section{}\label{App}

For completeness we also studied two more models for the $F_{\pi^0\gamma^*\gamma^*}(q^2_1,q^2_2)$. The first is based on the Regge theory and the second on the light-front holographic QCD. After generating a set of zero error data points for each model we fit the data with $P^L_1(Q^2)$ and $T^L_1(Q^2)$ sequences.

\subsection{Regge-model}

We consider first a Regge model based on the large-$N_c$ limit, $N_c$ been the number of colors, (see, for example, \cite{Arriola2006,Arriola2010,Kampf:2011ty,Czyz:2012nq} were similar large-$N_c$ models are used to fit directly the available data). In this limit, the vacuum sector of QCD becomes a theory of infinitely many non-interacting mesons and the propagators of the hadronic amplitudes are saturated by infinitely many sharp meson states. In the particular case below, the pion couples first to a pair of vector mesons $V_{\rho}$ and $V_{\omega}$ which then transform into photons. Thus we have:

\begin{align}\label{Reggemodel1}
&F_{\pi^0\gamma^*\gamma^*}(q^2_1,q^2_2)=\nonumber\\
&\sum_{V_{\rho},V_{\omega}}\frac{F_{V_{\rho}}(q^2_1)F_{V_{\omega}}(q^2_2)G_{\pi V_{\rho} V_{\omega}}(q^2_1,q^2_2)}{(q^2_1-M^2_{V_{\rho}})(q^2_2-M^2_{V_{\omega}})}+(q_1 \leftrightarrow q_2)\, ,
\end{align}
\noindent
where $F_{V_{\rho}}$ and $F_{V_{\omega}}$ are the current-vector meson couplings and $G_{\pi V_{\rho} V_{\omega}}$ is the coupling of two vector meson to the pion. The dependence on the resonance excitation number $n$ is the following
\begin{equation}
M^2_{V_{\rho}}=M^2_{V_{\omega}}=M^2+ n \Lambda^2\, ,\, \text{ and  }\,  F_{V_{\rho}}=N_c {V_{\omega}}\equiv F\, .
\end{equation}
\noindent
The combination of sums in Eq.~(\ref{Reggemodel1}) can be expressed in terms of the Digamma function $\psi(z)=\frac{d}{dz}log\Gamma(z)$:

\begin{eqnarray}\label{Reggemodel2}
\lefteqn{F_{\pi^0\gamma^*\gamma^*}(q^2_1,q^2_2) =  F_{\pi^0\gamma^*\gamma^*}(Q^2,A)=}\\ \nonumber
& &\frac{c}{N_c A Q^2} \left[ \psi\left(\frac{M^2}{\Lambda^2}+\frac{Q^2(1+A)}{2 \Lambda^2}\right)-\psi\left(\frac{M^2}{\Lambda^2}+\frac{Q^2(1-A)}{2 \Lambda^2}\right)\right]\, ,
\end{eqnarray}
\noindent
where $Q^2=-(q^2_1+q^2_2)$, $A=\frac{q^2_1-q^2_2}{q^2_1+q^2_2}$ and $c$ a constant.

To reassemble the physical case we consider $N_c=3$, $\Lambda^2=1.3$GeV$^2$ (as suggested by the recent light non-strange $q\bar{q}$ meson spectrum analysis~\cite{Regge_PM}), $A=1$ (which means $q^2_2=0$), $M^2=(0.8)^2$GeV$^2$ and the constant $c$ in such a way that the anomaly $F_{\pi^0\gamma\gamma}(0,0)=\frac{1}{4\pi^2f_{\pi}}$ is recovered.

Eqs.~(\ref{Reggemodel1}) and (\ref{Reggemodel2}) use the large-$N_c$ and chiral limits and thus have an analytic structure in the complex momentum plane which consists of an infinity of isolated poles but no brunch-cut (as does have the Log-model of section \ref{sec:model}), i.e. they become meromorphic functions. As such, they have a well-defined series expansion in powers of momentum around the origin with a finite radius of convergence given by the first resonance mass. It is well-known~\cite{Pommerenke} and largely explored in the context of Large-$N_c$~\cite{Masjuan,MasjuanThesis} than the convergence of any near diagonal PA sequence to the original function for any finite momentum, over the whole complex plane (except perhaps in a zero-area set) is guaranteed.

%The poles of the original Green’s function (i.e. the resonance masses) belong to this zero-area set because not even the original function is defined there, but there are also extra poles. These extra poles are called “defects” in the mathematical literature

For meromorphic functions such as Eqs.~(\ref{Reggemodel1},\ref{Reggemodel2}), another important result of Pad\'e Theory applies here, the Montessus de Ballore's theorem~\cite{Montessus,Baker}, which states that given a certain analytic function $f(z)$ at the origin which is meromorphic with exact $M$ poles in a certain disk on the complex plain, the sequence of PA converges uniformly to $f(z)$. In practice, provided $M$ is known ($M=1$ in our case), the Montessus' theorem asserts convergence for the sequence of M-pole Pad\'{e} Approximants $P_M^L$. These convergence theorems are confirmed by the good results collected on Table~\ref{TableRegge}, where after generating a set of zero-error data points with the model of Eq.~(\ref{Reggemodel2}), we fit them with the PA sequence and obtain the predictions for the $a_i$ coefficients.

With the PA $P^5_1$ we obtain a relative error of $0.02\%$, $2.9\%$, and $9.4\%$ for $a_0, a_1$ and $a_2$ respectively. The inclusion of the feasibility study at BES-III \cite{BESIII} decreases the error down to $2.4\%$ and $7.9\%$ for $a_1$ and $a_2$ resp. With a PTA sequence the results return $0.02\%$, $0.7\%$, and $0.8\%$ for $a_0, a_1$ and $a_2$ respectively when the PTA pole is located at $s_p=0.70$GeV$^2$

\begin{table}
\centering
\renewcommand{\arraystretch}{1.5}
{\scriptsize
\begin{tabular}{|c||c|c|c|c|c|c||c|}
\hline
 & $P^0_1$ &$P^1_1$ &$P^2_1$ & $P^3_1$ & $P^4_1$ & $P^5_1$ &  $F_{\pi^0\gamma^*\gamma}$ (exact) \\
\hline
$a_0 (GeV^{-1})$ & 0.2672 & 0.2730 &  0.2746 & 0.2751 & 0.2752 & 0.2753 &  0.2753 \\
$a_1 (GeV^{-3})$ & 0.2662 & 0.3121 &  0.3338 & 0.3457 & 0.3529 & 0.3571 &  0.3678 \\
$a_2 (GeV^{-5})$ & 0.2652 & 0.3600 &  0.4244 & 0.4616 & 0.4868 & 0.5030 &  0.5550\\
\hline
$\sqrt{s_p} (GeV)$& 1.00 & 0.92  &  0.87  & 0.86  & 0.85  & 0.84  & 0.80\\
\hline
\end{tabular}
}
\caption{$a_0,a_1$ and $a_2$ low-energy coefficients of the Regge-model in Eq.~(\ref{Reggemodel2}) fitted with a $P^L_1(Q^2)$
and its exact values (last column). We also include the prediction for the pole of each $P^L_1(Q^2)$ ($s_p$) to be compared with the lowest-lying meson in the model.}\label{TableRegge}
\end{table}

\begin{table}
\centering
\renewcommand{\arraystretch}{1.5}
{\scriptsize
\begin{tabular}{|c||c|c|c|c|c|c||c|}
\hline
 & $P^0_1$ &$P^1_1$ &$P^2_1$ & $P^3_1$ & $P^4_1$ & $P^5_1$ &  $F_{\pi^0\gamma^*\gamma}$ (exact) \\
\hline
$a_0 (GeV^{-1})$ & 0.2791 & 0.2774 &  0.2764 & 0.2759 & 0.2756 & 0.2754 &  0.2753 \\
$a_1 (GeV^{-3})$ & 0.3571 & 0.3362 &  0.3213 & 0.3108 & 0.3033 & 0.2986 &  0.2856 \\
$a_2 (GeV^{-5})$ & 0.4567 & 0.4031 &  0.3643 & 0.3358 & 0.3148 & 0.3009 &  0.2535 \\
\hline
$\sqrt{s_p} (GeV)$& 0.88 & 0.91  &  0.94  & 0.96  & 0.98  & 1.00  & 1.16\\
\hline
\end{tabular}
}
\caption{$a_0,a_1$ and $a_2$ low-energy coefficients of the Holographic-model in Eq.~(\ref{LFmodel}) fitted with a $P^L_1(Q^2)$
and its exact values (last column). We also include the prediction for the pole of each $P^L_1(Q^2)$ ($s_p$) to be compared with the lowest-lying meson in the model.}\label{TableLF}
\end{table}

\subsection{Holographic-model}

Finally, as a third model be analyze a simple holographic confining model presented in~\cite{BrodskyLF} (and also explored in Refs.~\cite{Mikhailov:2009kf,Agaev:2010aq,Brodsky:2011yv,Bakulev:2011rp}), based on light-front holographic QCD where the correct small $Q^2$ behavior (in order to simulate confinement) is introduced using the dressed current (see~\cite{BrodskyLF} for details)\footnote{We do not consider higher-twist components to keep the model easy to use.}.

In this context, the TFF is defined as

\begin{equation}\label{LFmodel}
F_{ \pi^0 \gamma^*\gamma}(Q^2)=\frac{P_{q\bar{q}}}{\pi^2 f_{\pi}} \int_0^1 \frac{dx}{(1+x)^2} x^{Q^2 P_{q\bar{q}}/(8\pi^2 f_{\pi}^2)}\, ,
\end{equation}
\noindent
where $P_{q\bar{q}}$ is the probability of finding the $q\bar{q}$ component in the pion light-front wave function. To reproduce the anomaly $F_{ \pi^0\gamma\gamma}(0)=1/(4 \pi^2 f_{\pi})$, we impose $P_{q\bar{q}}=0.5$.

This model reproduces quite well the transition form factor data up to $10$GeV$^2$ but disagrees in particular with BABAR's large $Q^2$ data (although compatible with Belle data within errors), specially because the model is reaching its asymptotic prediction ($Q^2 F_{\pi^0\gamma^*\gamma}(Q^2\rightarrow \infty)=2 f_{\pi}$~\cite{BL}) already at this medium-$Q^2$ region. Another interesting feature of this model is that no convergence theorem from Pad\'e Theory is known for this kind of function and then the Pad\'e convergence is not guaranteed in advance (in contrast to the previous Regge-model). It represents a robustness test of our method.

After generating again a set of zero-error data points with the model of Eq.~(\ref{LFmodel}), we use the PA sequence to fit these data and to obtain again the predictions for the $a_i$ coefficients. We collect the results in Table~\ref{TableLF}.\\

With the PA $P^5_1$ we obtain $0.04\%$, $4.6\%$, and $18.7\%$ as a relative errors for $a_0$, $a_1$ and $a_2$ respectively. With the inclusion of the feasibility study at BES-III \cite{BESIII}, we go down to $4.3\%$, and $17.1\%$ for $a_1,a_2$ resp. With the PTA sequence (the approximant pole located at $s_p=1$GeV$^2$) we obtain $0.6\%$, $4.8\%$, and $19.2\%$, respectively. Although no convergence theorem for this kind of function in Eq.~(\ref{LFmodel}) is known, the convergence of our PA sequence is clear. That is one of the most interesting features of the PA methods which is the convergence may occurs beyond expectations.\\

\end{document}